\documentclass[letterpaper, 10 pt, conference]{ieeeconf}  

\IEEEoverridecommandlockouts                              
\pdfoutput=1

\usepackage{graphicx} 
\usepackage{amsmath} 
\usepackage{amssymb}  
\usepackage[usenames, dvipsnames]{color}
\usepackage{url}
\usepackage{siunitx}
\usepackage{lipsum}
\usepackage{booktabs}
\usepackage{multirow}
\usepackage{subcaption}
\usepackage{float}
\usepackage{cite}

\usepackage[normalem]{ulem}

\usepackage{caption}
\usepackage[acronym]{glossaries}

\definecolor{darkpastelgreen}{rgb}{0.01, 0.75, 0.24}

\title{\LARGE \bf
Follow the Curve: Robotic-Ultrasound Navigation with Learning Based Localization of Spinous Processes for Scoliosis Assessment%
}

\author{Maria Victorova, Michael Ka-Shing Lee,
David Navarro-Alarcon and 
Yongping Zheng%
\thanks{This research was funded by the The Hong Kong Polytechnic University through the Intra-Faculty Interdisciplinary Project under grant ZVVR. \textit{Corresponding author: Maria Victorova.}}
\thanks{All authors are with The Hong Kong Polytechnic University (PolyU), Hung Hom, Kowloon, Hong Kong. D. Navarro-Alarcon and Y. P. Zheng are also with the Research Institute for Smart Ageing, PolyU, Hong Kong. (e-mail: maria.victorova@connect.polyu.hk)}
}

\makeglossaries
\newglossaryentry{maths}
{
        name=mathematics,
        description={Mathematics is what mathematicians do}
}

\newglossaryentry{formula}
{
        name=formula,
        description={A mathematical expression}
}

\newacronym{us}{US}{ultrasound}
\newacronym{DOF}{DOF}{degree of freedom}
\newacronym{UR}{UR}{universal robot}
\newacronym{msd}{MSD}{musculoskeletal disorders}
\newacronym{DL}{DL}{deep learning}
\newacronym{AIS}{AIS}{adolescent idiopathic scoliosis}
\newacronym{GUI}{GUI}{graphical user Interface}
\newacronym{UR5}{UR5}{universal robot model 5}
\newacronym{API}{API}{application programming interface}
\newacronym{SD}{SD}{standard deviation}
\newacronym{MSE}{MSE}{mean squared error}
\newacronym{CNN}{CNN}{convolutional neural network}
\newacronym{FCN}{FCN}{fully convolutional network}
\newacronym{BMI}{BMI}{body mass index}
\newacronym{SPA}{SPA}{spinous process angle}

\renewcommand{\baselinestretch}{0.99}

\begin{document}

\maketitle
\thispagestyle{empty}
\pagestyle{plain}

\begin{abstract}

The scoliosis progression in adolescents requires close monitoring to timely take treatment measures. Ultrasound imaging is a radiation-free and low-cost alternative in scoliosis assessment to X-rays, which are typically used in clinical practice. 
However, ultrasound images are prone to speckle noises, making it challenging for sonographers to detect bony features and follow the spine's curvature. 
This paper introduces a robotic-ultrasound approach for spinal curvature tracking and automatic navigation. 
A fully connected network with deconvolutional heads is developed to locate the spinous process efficiently with real-time ultrasound images. 
We use this machine learning-based method to guide the motion of the robot-held ultrasound probe and follow the spinal curvature while capturing ultrasound images and correspondent position. 
We developed a new force-driven controller that automatically adjusts the probe's pose relative to the skin surface to ensure a good acoustic coupling between the probe and skin. 
After the scanning, the acquired data is used to reconstruct the coronal spinal image, where the deformity of the scoliosis spine can be assessed and measured. 
To evaluate the performance of our methodology, we conducted an experimental study with human subjects where the deviations from the image center during the robotized procedure are compared to that obtained from manual scanning. The angles of spinal deformity measured on spinal reconstruction images were similar for both methods, implying that they equally reflect human anatomy.
 
\end{abstract}


\section{Introduction}\label{sec:introduction}
While ultrasound (US) has been proven to be a safe and reliable technique for scoliosis assessment \cite{Zheng2016}, it is difficult for operators to identify anatomical features in ultrasound images due to its inherent speckle noises. 
The correct detection of bone features is crucial to properly follow the spinal curvature of scoliosis patients during the scanning. 
Thus, the quality of the resulting 3D spinal reconstruction is highly dependant on the sonographer's experience. 

\begin{figure}[]
    \centering
    \includegraphics[width=\linewidth]{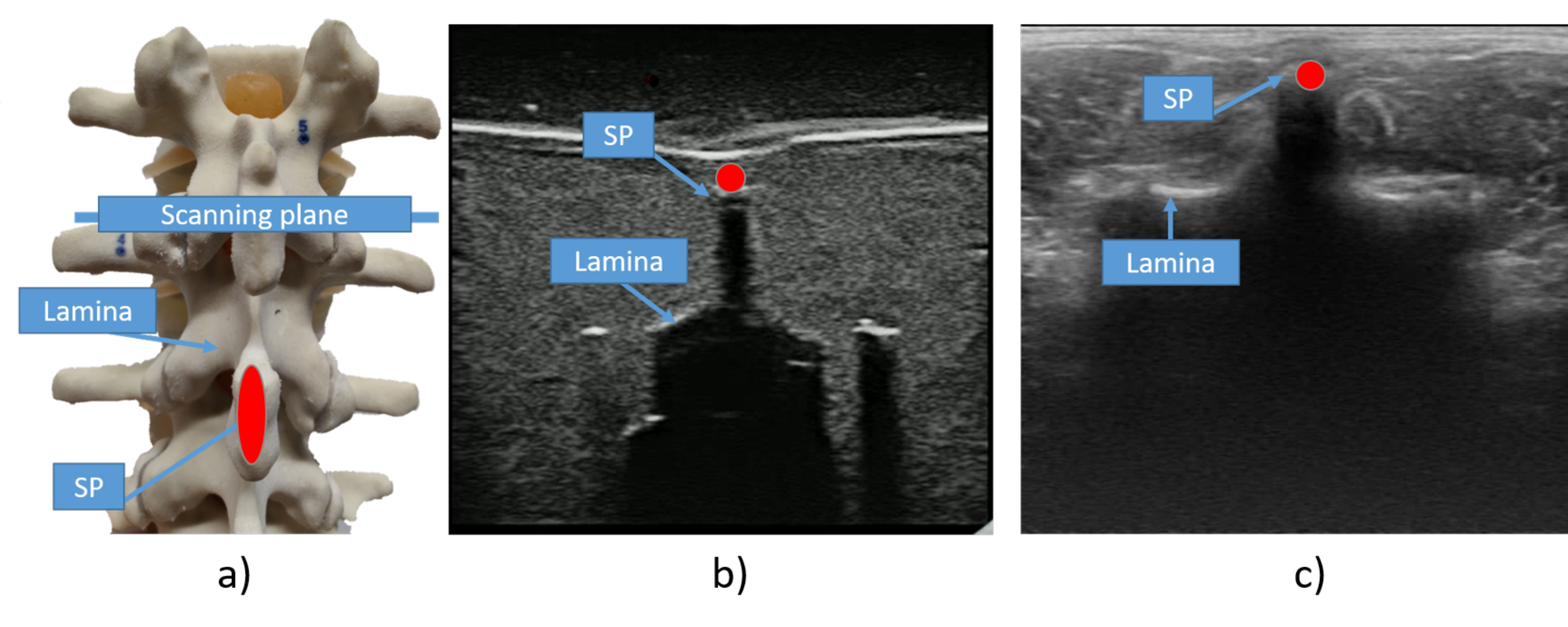}
    \caption[Spinal anatomical features on phantom and US images of phantom and human]{Spinal anatomical features on phantom and US images of phantom and human. Lamina and spinous process (SP). a) Spinal features presented at the spinal column phantom, b) Spinal features visible on US images of a phantom, c) Spinal features visible on US images of a human.}
    
    \label{fig:spinal_features}
\end{figure}

This limitation can be overcome by using computer control to locate these features and guide a robotic arm that manipulates an ultrasound probe.
To ensure that the spine is always in the probe's field of view, the robot needs to follow the spine's profile during scanning so that the vertebrae are located in the center of the ultrasound image. 
A spinous process is located in the middle of the vertebrae and indicates the vertebra's presence in the image, whereas its absence indicates an intervertebral gap (see Fig. \ref{fig:spinal_features}).
These features are utilized for 3D spine reconstruction and scoliosis assessment by measuring the angle of the spinal curvature. 

There is a number of researchers that have used Phase Symmetry (PS) as a set of filters to enhance bony structures, which can be used in image processing techniques to locate spinal features \cite{PurangAbolmaesumiSeptimiuE.SalcudeanWen-HongZhuMohammadRezaSirouspour2002, Hacihaliloglu2009}. 
Other work \cite{Berton2016b} has combined PS with machine learning (e.g., linear discriminant analysis classifier) to segment anatomical features in a US image frame as spinous processes, acoustic shadows, and other tissues. 
Another method \cite{villa2018fcn} have used \acrfull{FCN} together with phase symmetry to segment the bone surface in a US image. 
The input image to this FCN has three channels, where the first channel is an original image, the second is the PS-processed image, and the third is the image resulting from computing the confidence map \cite{karamalis2012ultrasound}. 
Methods based on pure deep learning techniques are few and are mainly focused on detecting the spine in different probe orientations (e.g., sagittal); These are typically used for spinal injections where the features appear different and are not suitable for scoliosis assessment \cite{Hetherington2017,Baka2017}.

There are various limitations with the above state-of-the-art methods, e.g., the considerable uncertainty in the estimation where is the intervertebral gap; the need to segment pixels containing bony structures manually; multiple parameters that must be tuned in all three different algorithms (i.e., PS, confidence map, and FCN). 
Therefore, these methods are not suitable for continuous spinal scanning.

There is currently no algorithm designed explicitly for spinous process localization (representing the center of the vertebrae) that we can use to track the spine in our scoliosis application.
Therefore, we take inspiration from the area of human pose estimation, which also requires accurate landmarks localization in the form of joints. 
One method frequently used in this field is landmark pose regression from heatmaps generated by FCN, where each pixel represents a probability of containing the sought landmark. 
Various researchers have adopted this approach for localization of these types of landmarks \cite{newell2016stacked, xiao2018simple,Nibali2018}.

The original contribution of this study is a spinal curvature following approach with intelligent robot-driven ultrasound for scoliosis assessment.
Our developed prototype is guided by real-time ultrasound images and has spinal curvature following capabilities. 
To accurately detect bone features from ultrasound images, our method uses a robust deep learning based network specifically designed for spinous process localization. 
The network takes as input raw ultrasound images, and it outputs a spatial heatmap that indicates the location probability of the spinous process; The maximum probability predicts the position of these bony features on each ultrasound image.
The predicted location enables the system to guide the robotic probe along the curved spine by using an automatic motion controller combined with a Kalman filter estimator, which enables forming a continuous spinal path compensating for the intervertebral gaps. 
The proposed robotic scanning approach was evaluated on both spinal phantoms and human volunteers; Its performance was compared with the conventional manual scanning approach. 

The rest of this manuscript is organized as follows: Sec. II presents the methodology. Sec. III reports the conducted experiments. Sec. IV gives final conclusions.

\section{Methods} \label{sec:methods}
\subsection{Robotic-ultrasound scoliosis examination procedure}
\label{sec:examination_procedure}
During the regular ultrasound scoliosis assessment developed by Zheng et al., \cite{Zheng2016} sonographer manually scans the human spine with an ultrasound probe in caudo-cranial direction, trying to center the spine in the field of view of the probe. The sonographer has to apply pressure during the scan and rotate it to ensure that the probe maintains tight contact with the patient's skin. The ultrasound system used for manual scanning is Scolioscan Air \cite{scolioAirlai2021}, which consists of a USB Ultrasound probe and a tablet with Scolioscan Air software, which receives images and coordinates of the probe for spinal 3D reconstruction.

The robotic scanning process\footnote{Ethical approval HSEARS20210417002 was given by Departmental Research Committee (on behalf of PolyU Institutional Review Board)}, (the setup is shown in Fig. \ref{fig:method_and_setup_human} b)), begins by manually positioning the robotic arm operated ultrasound probe at the sacrum level; from there, the robotic arm travels towards the subject's back until the force meets the force setpoint. The probe then begins to go higher along the subject's back.
The robot provides continual pressure to the subject's back during the motion and follows the spinous process as a reference characteristic of the spine's center. The robotic procedure uses a pitch rotation to keep the probe normal to the subject's back and a yaw rotation to make sure the left and right sides of the probe surface are in contact with the skin.
After both manual and robotic procedures, the software generates coronal images as slices of the 3D spinal reconstruction for curvature measurement, according to the spinous process angle method presented in \cite{Zheng2016}. The angle is calculated between the most tilted vertebrae of the scoliotic curve. During both scanning approaches, human subjects could breathe normally, and the posture was fixed by asking the subject to lean on the built-in supporters as in Fig. \ref{fig:method_and_setup_human} b).
 


\subsection{Materials and experimental setup}
\begin{figure}[]
    \centering
    \includegraphics[width=\linewidth]{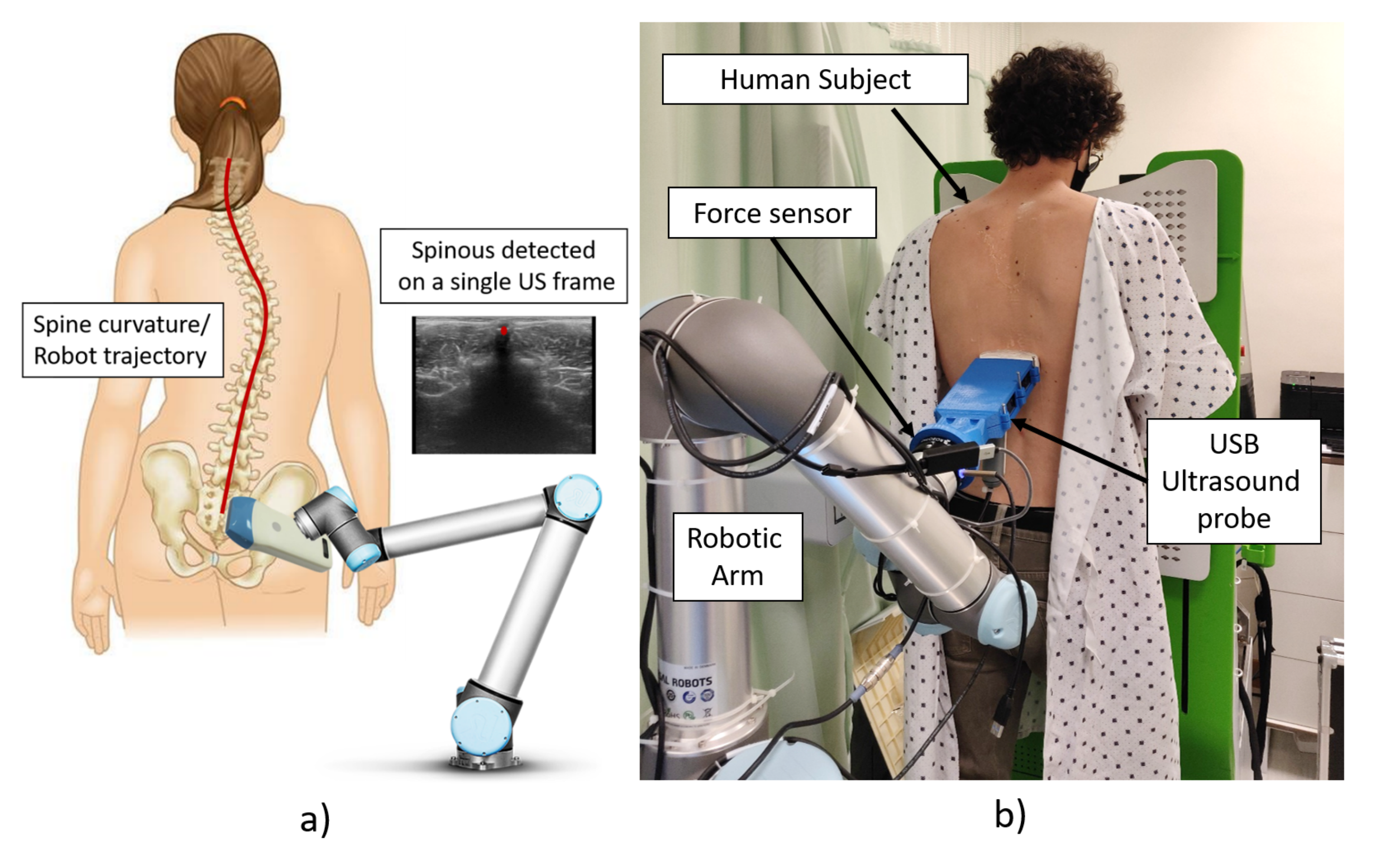}
    \caption[Setup for robotic scoliosis assessment for human scanning]{a) Proposed spinal curvature tracking. The robotic arm with ultrasound probe follows the spinal curvature by real-time detection of spinous processes on ultrasound frames. b) Setup for robotic scoliosis assessment for human scanning. Robotic arm UR5 presented with mounted force sensor FT300 and ultrasound probe in 3D printed probe holder. The human subject is stabilized by the supporters mounted on the frontal plate of the Scolioscan machine.}
    \label{fig:method_and_setup_human}
\end{figure}

Fig. \ref{fig:method_and_setup_human} b) shows the setup for automatic scoliosis assessment using a robotic arm - (Universal robot UR5) with force sensor mounted (FT300, Robotiq) and USB ultrasound probe (Sonoptek, Beijing).
The Ultrasound probe (Fig. \ref{fig:setup_coordinate_systems} b)) captures images with 7.5 MHz, depth of 6 cm, and sends raw data at 10 fps over USB to PC, where the images are formed in a size of 640x480 pixels. The probe's aperture is rectangular with a more extended size of 80mm.
The robot was connected via TCP/IP protocol with a PC, where the control algorithm was launched at a rate of 30 fps Hz. 
Since the robotic control has a higher frequency than the ultrasound images frame rate, several robotic positions correspond to a single image.


\subsection{Coordinate systems}

\begin{figure}[]
    \centering
    \includegraphics[width=0.9\linewidth]{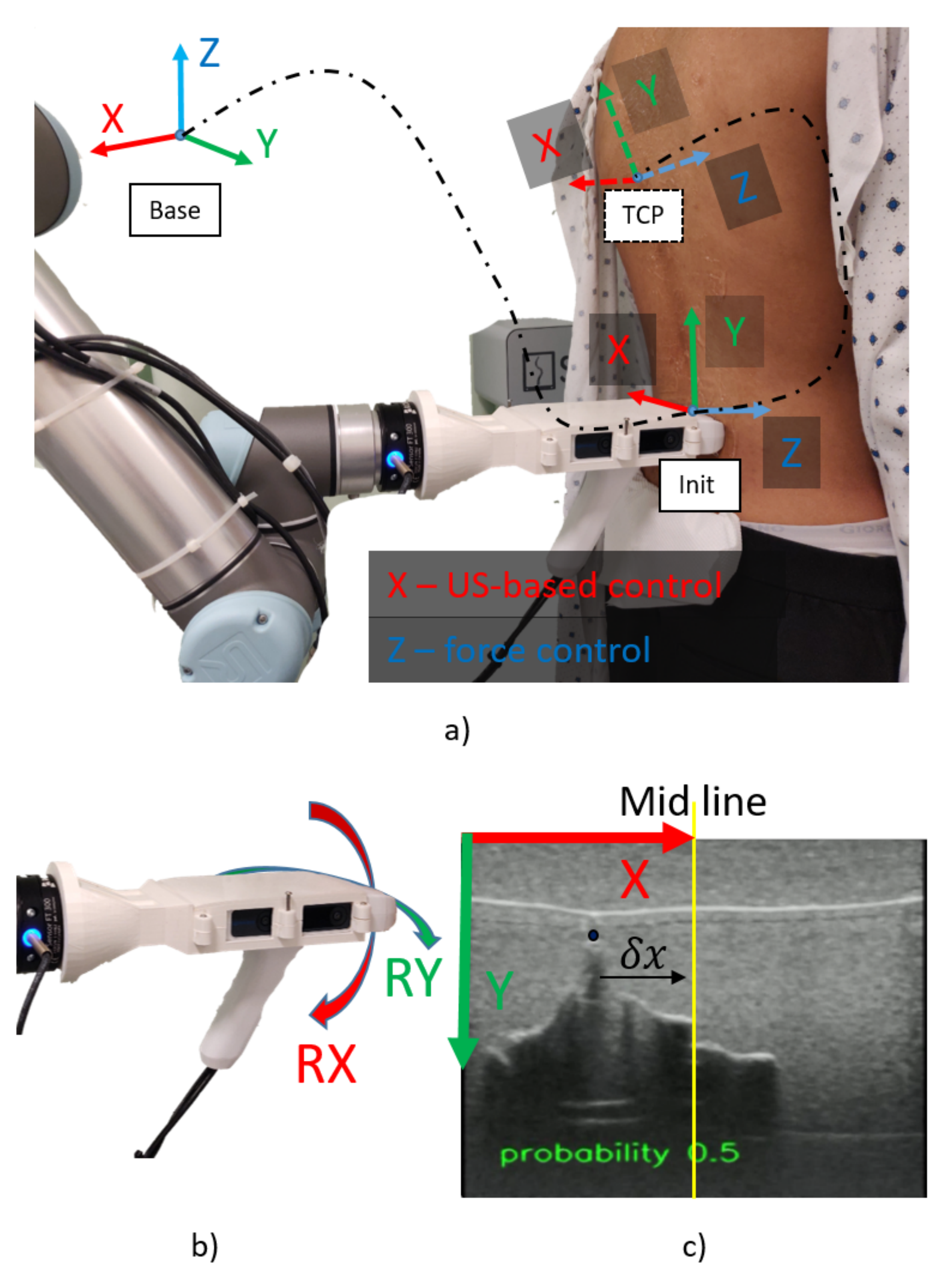}
    \caption[The robotic-US setup with coordinate frames labelled]{a) The robotic-US setup with coordinate frames labelled.
    b) Ultrasound Probe 3D-printed holder. c) B-mode US image coordinate frames.}
    \label{fig:setup_coordinate_systems}
\end{figure}

Fig. \ref{fig:setup_coordinate_systems} a) shows different coordinate systems for robotic-US setup. The Base coordinate system is fixed for all times with the origin at the point where the robot is attached to the working table. 
The Init coordinate system is initialized upon the start of each scanning procedure, placing the US probe perpendicular to the subject's back and parallel to the ground to match the settings of manual scanning. Tool Central Point (TCP) coordinate frame coincides with the Init frame upon start, but as the US probe moves along the subject's back, the TCP frame moves together with the US probe. The TCP frame is used for control purposes.
In Fig. \ref{fig:setup_coordinate_systems} b) the rotations of the probe in TCP coordinate frame are presented. 
The ultrasound images have two axes, X is along the image width, and Y is along with the height, and, at the same time, it represents the penetration depth in spinal tissues. The image has a resolution of 640x480 pixels.


\subsection{Dataset}

The acquired data set consists of 24 scoliosis patients, which were scanned manually by a sonographer. Each ultrasound frame for each subject was manually labeled by an ultrasound expert indicating the spinous process location, if present, as shown in Fig. \ref{fig:spinal_features} c). The target heatmap images were generated by applying a 2D Gaussian distribution aligned with the manual label's center. The resulting dataset consists of spinous process images (not including intervertebral images, where the spinous process is absent) and correspondent target heatmap images. The training subset has 19 subjects, 9,396 spinous process images, and the test has five subjects of 2,972 spinous process images.

\subsection{Spinal features localization}

\begin{figure}[]
    \centering
    \includegraphics[width=\linewidth]{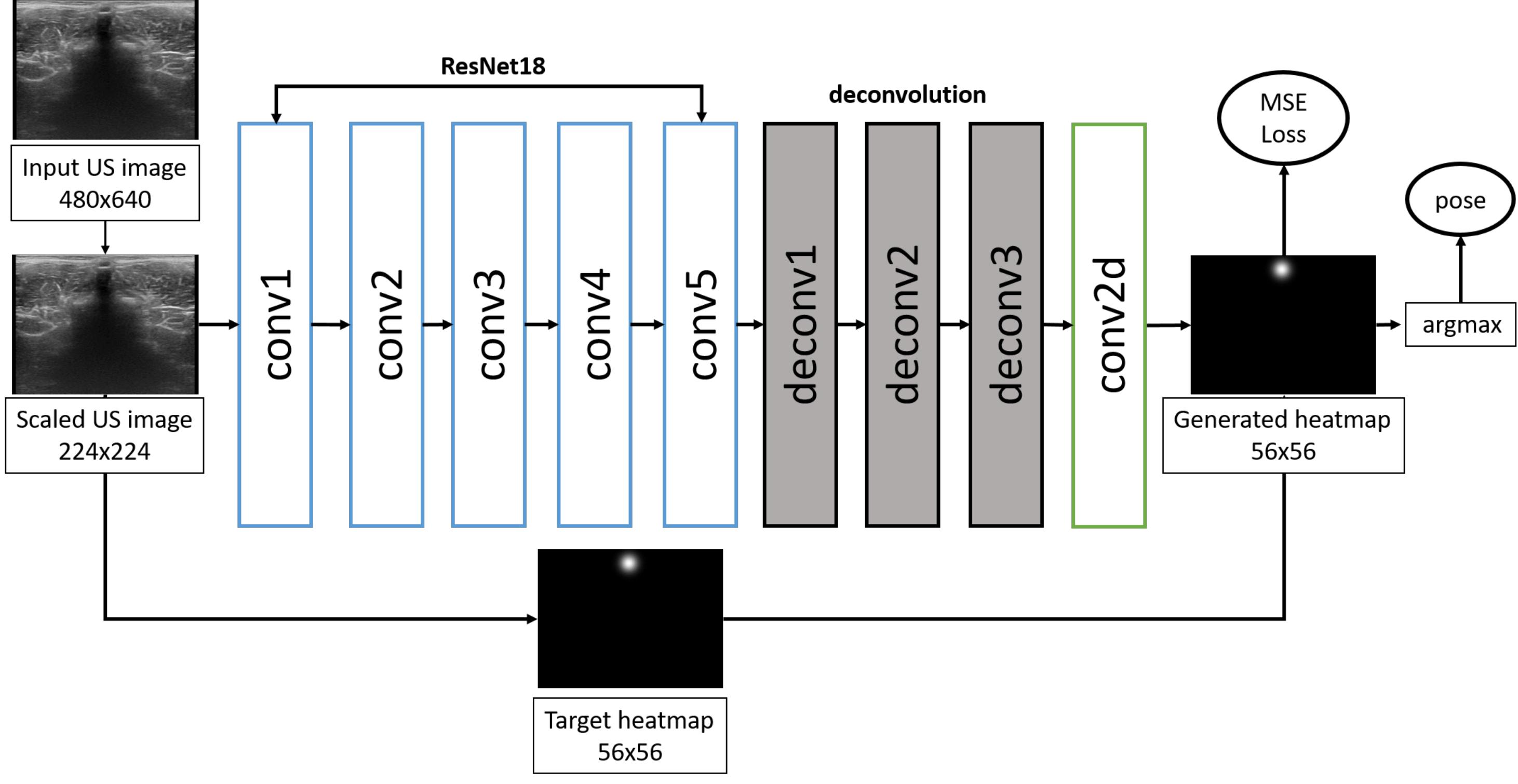}
    \caption[Proposed spinous localization network]{Proposed spinous localization network. ResNet based fully connected network with deconvolutional head. Loss is calculated as mean squared error between predicted and target heatmaps. The final pose is calculated from max intensity of the predicted heatmap.}
    \label{fig:FCN_diagram}
\end{figure}

The method of spinous process localization uses the heatmap approach. The heatmap represents the probability of each pixel of the image to be a spinous process, forming the Gaussian distribution around the point with the maximum probability; that point is the sought spinous process location.
The schematic overview of our proposed network for spinous process classification and localization is conceptually shown in Fig. \ref{fig:FCN_diagram}. The widely used ResNet is used as a backbone to extract image features from input ultrasound images. The conv1 to conv5 layers in the figure represent the five convolutional stages in ResNet. 

We replaced the last fully-connected layer of the ResNet (which gives the class distribution score) with three deconvolutional layers with batch normalization \cite{ioffe2015batch} and ReLU activation \cite{krizhevsky2012imagenet}, which act as the decoder to generate the image features heatmap \cite{xiao2018simple}. Each deconvolutional layer has 256 filters with a 4x4 kernel; the stride is 2. The deconvolutional layers are followed by one 1x1 convolutional layer, which transforms the resulted features matrix to a final heatmap where each image pixel intensity represents the probability of being one of the classes; in our case, the maximum intensity represents the high probability of the pixel belonging to a spinous process class. The loss between the ground truth and predicted heatmap is calculated using the mean squared error (MSE). 
The final location is derived from the maximum intensity pixels of the resulted heatmap. To increase the robustness of the dataset, we use various image data augmentation techniques, such as rotation, horizontal and vertical flipping. 


\subsection{Spinal features tracking}
\begin{figure}[]
    \centering
    \includegraphics[width=\linewidth]{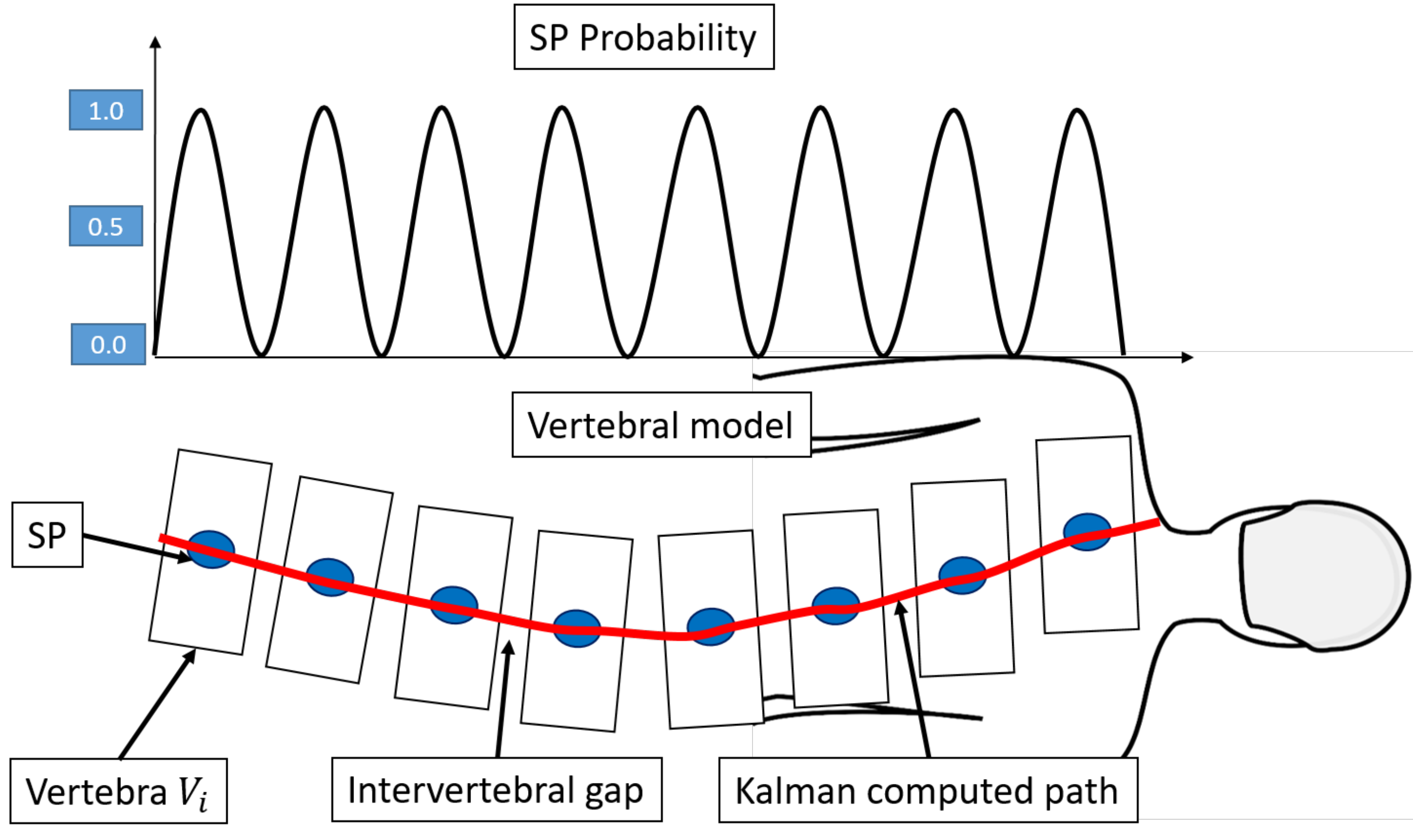}
    \caption{Model of vertebral and intervertebral gap distribution in the spine and correspondent expected network output.}
    \label{fig:Kalman_computed}
\end{figure}

The sequence of US images for one human subject is called ultrasound sweep. In the sweep, the spinal images of the spinous process alternate with intervertebral gap images (shown in Fig. \ref{fig:Kalman_computed}). 
Since the localization network does not provide meaningful location information for intervertebral gap (the detection probabilities are less than $50\%$ and thus the locations are rejected), there is a need for a method that can form a continuous path.
To overcome the issue with missing location information, we generate Kalman computed continuous spinal path. Kalman filter \cite{welch1995introduction} was used to fuse filter-predicted locations $\hat{x}_{k-1}$ with network detected locations $x_{net}$, whenever they are available, as expressed in eq. \ref{eq:Kalman}. Here $K_k$ is Kalman gain, and $\hat{x}_k$ is a resulted next-frame location. This method also helps to filter the network output to generate smoother locations as inputs for image-based robotic control.
\begin{equation}
 \hat{x}_k=\hat{x}_{k-1}+K_k(x_{net}-\hat{x}_{k-1})
    \label{eq:Kalman}
\end{equation}
During the ideal scanning procedure, the spinal path plot will be a straight line positioned at the center of the coordinates (half of the image width size), which corresponds to the scanning when the spinous process was kept at the center of the field of view of the ultrasound probe. Any deviations of the points in the plot from the center of the coordinates can be considered inaccuracies of the scanning procedure. 
 
\subsection{Learning based robotic navigation}

Ultrasound feature-based navigation control works along X-axis of TCP coordinate frame $v_{x}$. Each captured ultrasound frame used as an input for pre-trained deep learning model (Fig. \ref{fig:FCN_diagram}), which outputs the predicted location of spinous process and probability of this point belonging to the spinous process (confidence of prediction). The locations with probability higher than the threshold considered to be valid detected spinous process. Since the resulted heatmap is 56x56, the location output should be re-scaled to correspond to original image size of $w_{image} = 640, h_{image} = 480$. Those locations are fused with Kalman filter predicted locations generating the location coordinate $x_{image}$ for robotic control, expressed in coordinate frames as shown in Fig. \ref{fig:setup_coordinate_systems} c). The error between the spinous location and the image center $\delta x_{pixels} = x_{image}-w_{image}/2$, can be re-scaled to be expressed in meters, which is the same dimensions as the output velocity $\delta x = \frac{\delta x_{pixels}d_{probe}}{w_{image}}$.
The proportional control law using the spinous process location as feedback would look as following: $v_{x}=-K_{im}\delta x$, where $K_{im}$ is a proportional control gain. There are two different control gains used depending on the distance to the desired spinous location. This approach is made to prevent sudden velocities generated in reaction to far located points. Adding exponential smoothing with control gain $\alpha$ to the control law to ensure smoother velocity output would transform this equation to following:
\begin{equation}
 v_{x}(t) = \alpha(-K_{im}\delta x)+(1-\alpha)v_{x}(t-1)
    \label{eq:velocity_image_control}
\end{equation}

\subsection{Force-driven acoustic coupling}

The manipulation of US probes over the body involves two main problems: (1) stable interaction with deformable tissues; and (2) feature-based navigation control. The former is needed to ensure that the probe provides stable contact with human skin surface to get the US images of acceptable quality; The latter is needed to autonomously conduct the scanning procedure following the spinal curvature. Thus, a velocity control during the US-guided procedure uses mainly two parameters to form a control loop: force screw $f = (f_{x} \;f_{y} \;f_{z} \;m_{x} \;m_{y} \;m_{z})^T$ and spinous process location in each US images $X_{im}$. The control velocity vector is expressed as $v_{TCP} = (v_{x} \;v_{y} \;v_{z} \;r_{x} \;r_{y} \;r_{z})^T$.
During the scanning, the robot moves along the spine in the direction on the Y-axis of TCP coordinate frame with constant velocity $v_{y}$ set before scanning. 
PID velocity control is used for force control along Z-axis of TCP coordinate frame $v_{z}$ according to the Fig. \ref{fig:setup_coordinate_systems} similar to \cite{Victorova2019}.
While the traditional PID control works well for the applications where setpoint (in our case it is $F_{ref}$) is not changing, however when the setpoint changes, the sudden spike appears in the velocity output. The setpoint change necessity comes from an idea of changing force settings according to the spinal region (see section \ref{sec:spinal_regions}). One way to overcome this issue is to use ``derivative on measurement", which neglects the reference force change, $\frac{dF_{ref}}{dt} = 0$. 
Since $\frac{de}{dt} =\frac{dF_{ref}}{dt} - \frac{dF_{curr}}{dt}$, the resulted equation:
\begin{equation}
 v_{z} = K_{p}e + K_{i}\int{edt} -
 K_{d}\frac{dF_{curr}}{dt},
    \label{eq:PID_on_measurement}
\end{equation}

where $e = F_{ref} - F_{curr}$ is an error between reference force $F_{ref}$, set prior to scanning, and current force measurement $F_{curr} = f_{z}$, coming from force sensor in the Z direction. The control gains $K_{p},K_{i},K_{d}$ are chosen experimentally according to the intrinsic parameters, such as the robotic system response and extrinsic parameters, such as stiffness of the subject's back.

In addition to a PID control, the saturation $v_{lim}$ of the output is used as a safety measure to prevent larger velocities from affecting the stable contact with the skin surface or causing any discomfort for the subject during the scan. There is also a custom set limit for the maximum force exerted on the robot end-effector. If $F_{crit}$ is reached, the robot stops the operation to ensure the safety of the patient and operator.

\subsection{Force-based probe orientation adjustments}
\begin{figure}[]
    \centering
    \includegraphics[width=\linewidth]{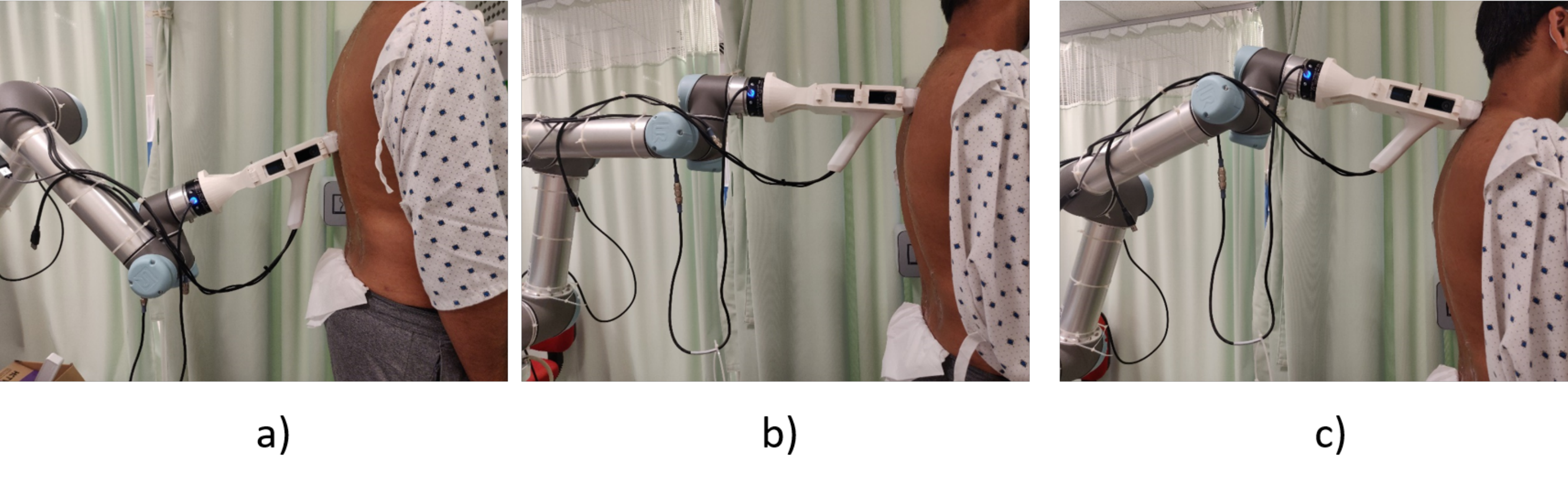}
    \caption[The probe fixed rotation adjustment based on the shape of the following surface.]{The probe fixed rotation adjustment based on the shape of the following surface. The probe orientation is maintained normal to the skin surface, the $r_x$ value is fixed.}
    \label{fig:tilt_force_human}
\end{figure}

To ensure the robotic actuated ultrasound probe smoothly adapts to the curvature change of the human back surface, the moment (torque) $m_{x}$ induced control is used. The example of the procedure with the force-compliant scanning is presented in Fig. \ref{fig:tilt_force_human}. The coordinate frames of the force sensor and TCP frame used to control coincide; thus, the $r_{x}$ rotation will affect readings of $m_{x}$. The $r_{x}$ rotation change during the scanning ensures that that the ultrasound probe faces the human back surface at an angle close to $90^{\circ}$. The proposed control follows the following law: $r_{x} = -K_{pitch}m_{x}$.

The principle behind this law is that the rotation $r_{x}$ will compensate the moment $m_{x}$ change during the scanning. The control gain $K_{pitch}$ specifies how sensitive it would be the rotation and how fast the compensation would occur.

\subsection{Spinal regions classification for scanning settings adjustments}
\label{sec:spinal_regions}

\begin{figure}[]
    \centering
    \includegraphics[width=\linewidth]{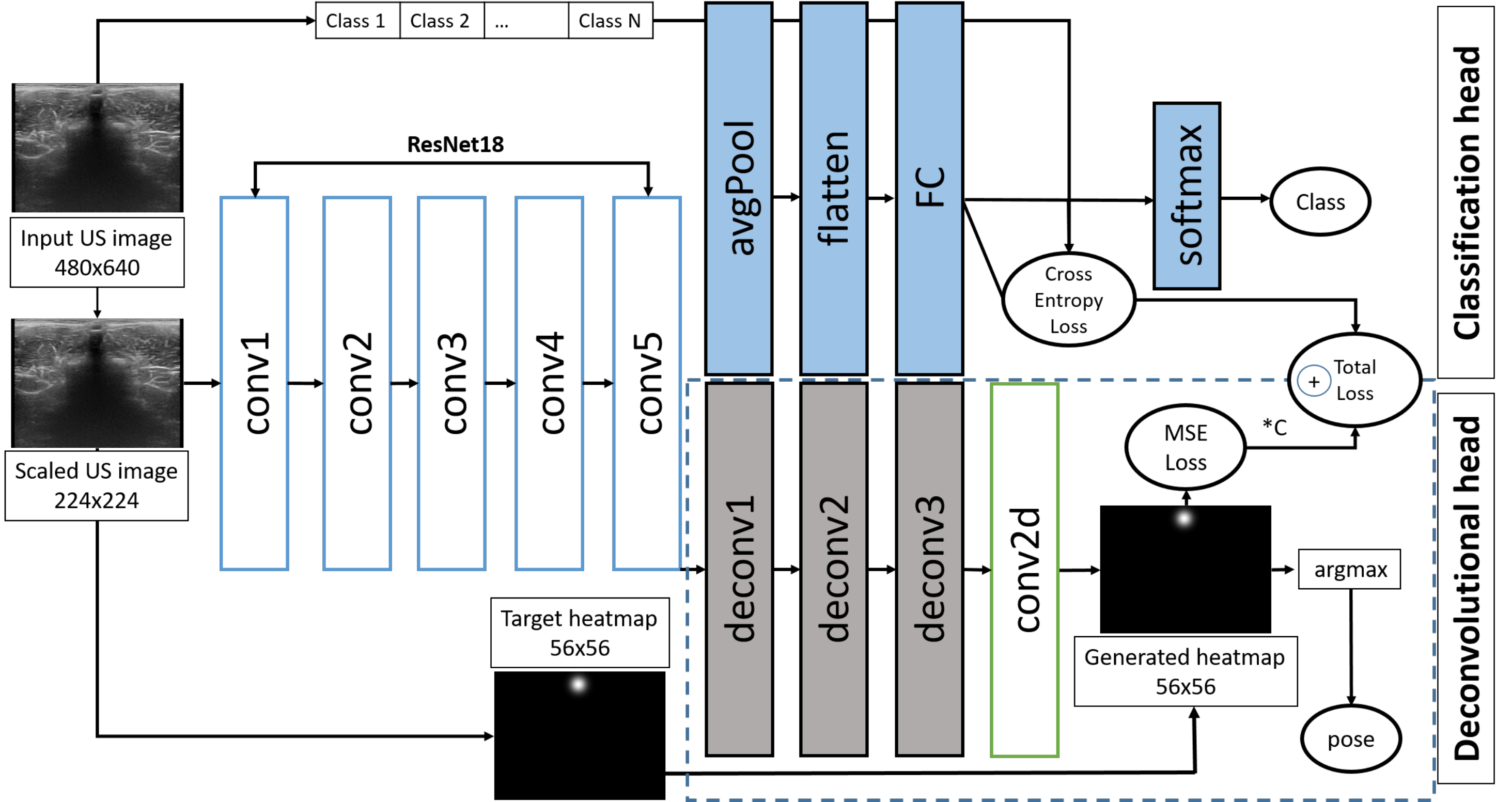}
    \caption[Network for Spinal region classification and spinous process localization tasks.]{Network for Spinal region classification and spinous process localization tasks.}
    \label{fig:region_net_diagram}
\end{figure}

Due to anatomical spine structure, the lumbar region of the spine usually lays deeper under the layer of the fat and muscles than the thoracic region of the spine. According to these anatomical differences, the pressure applied by the sonographer during the scanning to those regions is slightly different; the lumbar region of the spine requires higher forces applied to compress the layer of fat and capture the entire bony structures. 
We propose using the network for spinal regions classification to detect when the lumbar region ends and thoracic starts. The robotic scanning settings such as force applied and ultrasound settings as ultrasound gain could be adjusted by detecting this region change. 

The proposed network has two heads for two different tasks, classification and localization; the first is a deconvolutional head for localization task is similar to one presented in Fig. \ref{fig:FCN_diagram}; the second is a classification head used to output the probability of an image to belong to a certain spinal region. The diagram of resulted multi-task network is presented in Fig. \ref{fig:region_net_diagram}. The network is trained end-to-end, which means the total loss combined from two network heads is used for back-propagation. The total loss is a sum of two losses with a scaling factor, $C$, which is used to balance the magnitudes of the two losses, $Total Loss = Loss_{classification} + C*Loss_{localization}$.
The resulted class number where the image frame belongs can be found by applying a softmax layer on the probabilities output from the classification head. The loss for localization task is MSE and for classification is Cross-entropy loss. 


\begin{figure}[]
    \centering
    \includegraphics[width=\linewidth]{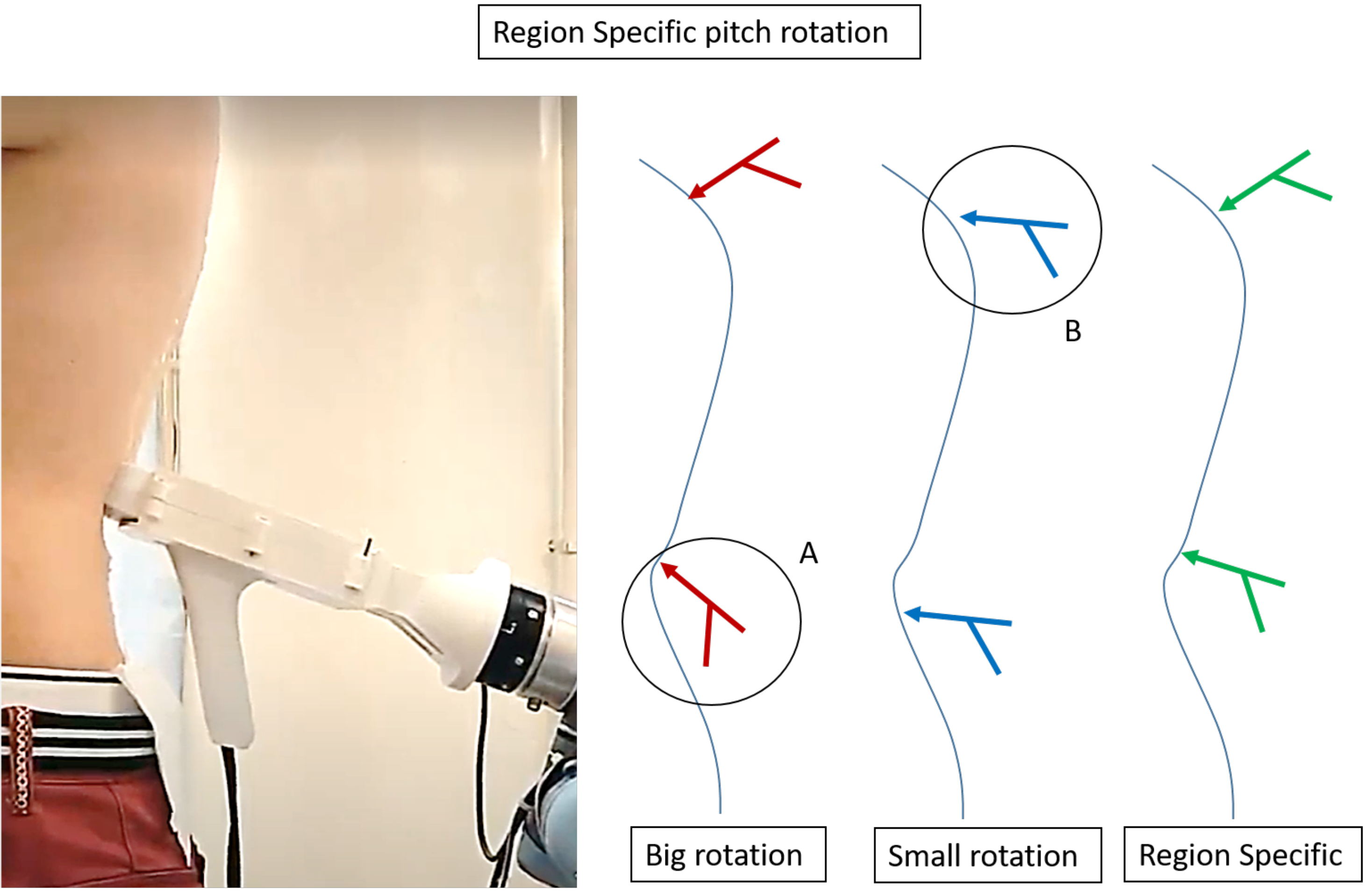}
    \caption[Region Specific Pitch Rotation]{Region Specific Rotation. The proposed application of spinal region classification network. The cases A and B indicates problems with fixed $r_x$ value selection. The region specific scanning proposes the change of $r_x$ value depending on spinal region.}
    \label{fig:region_rotation}
\end{figure}

Fig. \ref{fig:region_rotation} shows the proposed application for the spinal regions classification network. The probe in case A is too tilted towards the patient, which might cause the contact to lose and the uncomfortable feeling of the patient. The highly curved lumbar lordosis can cause this issue, which is quite widespread among teenagers. In this case, if the pitch rotation value $r_x$ is low, the probe can freely scan the lumbar region with minimal rotations. However, this potentially can lead to case B for the subjects with the curved upper back. If the value $r_x$ is low, the probe will not follow the surface correctly and eventually lose contact with the human back. The region-specific $r_x$ value change was tested to solve the issue with contact losing.

\section{Results}
\subsection{Spinal features localization method's results}

\begin{figure}[]
    \centering
    \includegraphics[width=\linewidth]{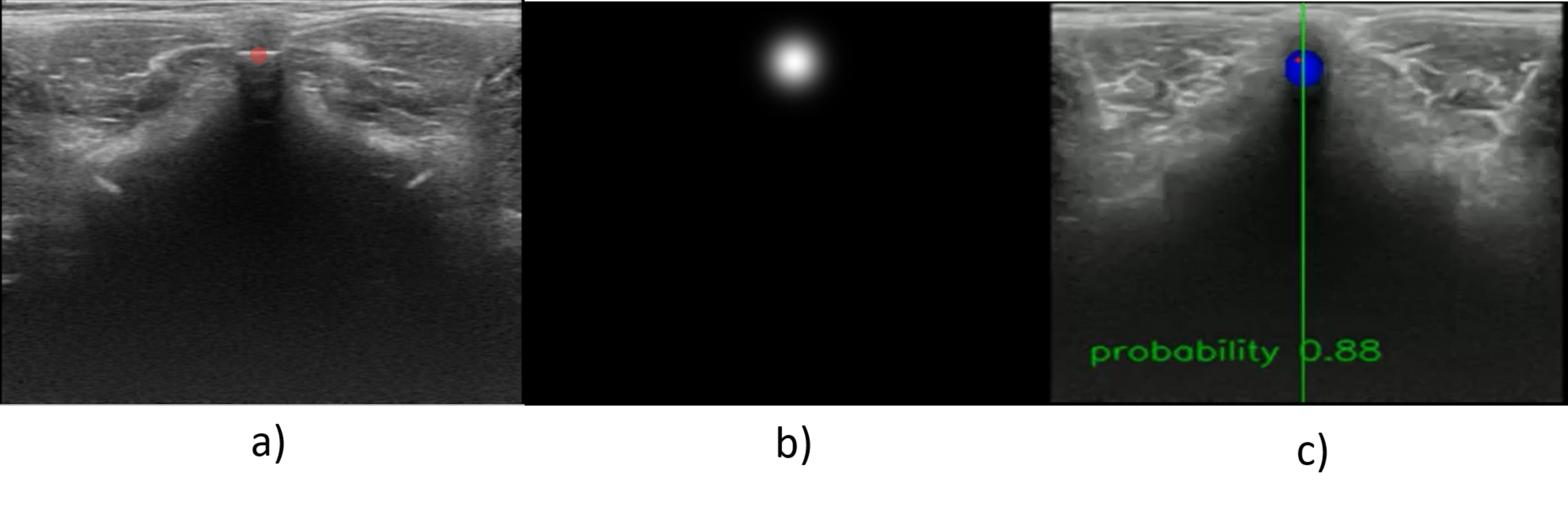}
    \caption[Results of the spinous process localization on a single frame]{Results of the spinous process localization on a single frame. a) The US frame manually labelled; b) Ground truth mask: the resulted target heatmap with a Gaussian drawn around manual label center; c) localized spinous process with proposed architecture (red dot), ground truth Gaussian (blue) and middle line of the image (green). The resulted confidence of the localized point is $88 \%$.}
    \label{fig:FCN_detection_method}
\end{figure}

The model presented in Fig. \ref{fig:FCN_diagram} was trained for 100 epochs reaching the best validation accuracy at the 10th epoch; learning rate was 0.0001, batch size is 12. Adam \cite{kingma2014adam} optimizer was used. The initial weights for ResNet \cite{he2015deep} backbone were loaded from publicly available ImageNet pre-trained ResNet18 model. The inputs for the model were ultrasound images of the spinous process collected from 19 subjects. The images were normalized (so that the pixels becoming floats in range [0, 1]) and resized to a size of 224x224 pixels. The targets were heatmaps generated from manually labelled images, as shown in Fig. \ref{fig:FCN_detection_method} a) and Fig. \ref{fig:FCN_detection_method} b). The resulted model was tested on spinous process images collected from 5 subjects (2971 images).

The accuracy of detected spinous process locations is calculated similarly to a PCK (percentage of correct key points) metric used in human pose estimation tasks. The detected spinous process is considered correct if the distance between the predicted and target spinous process is within a threshold, here $50 \%$. The resulting accuracy on a test set (2,972 SP images) was $97.8 \%$ with a mean distance error of 8 pixels, corresponding to 1.0mm.

\subsection{Spinal features tracking method's results}
\begin{figure}[]
    \centering
    \includegraphics[width=\linewidth]{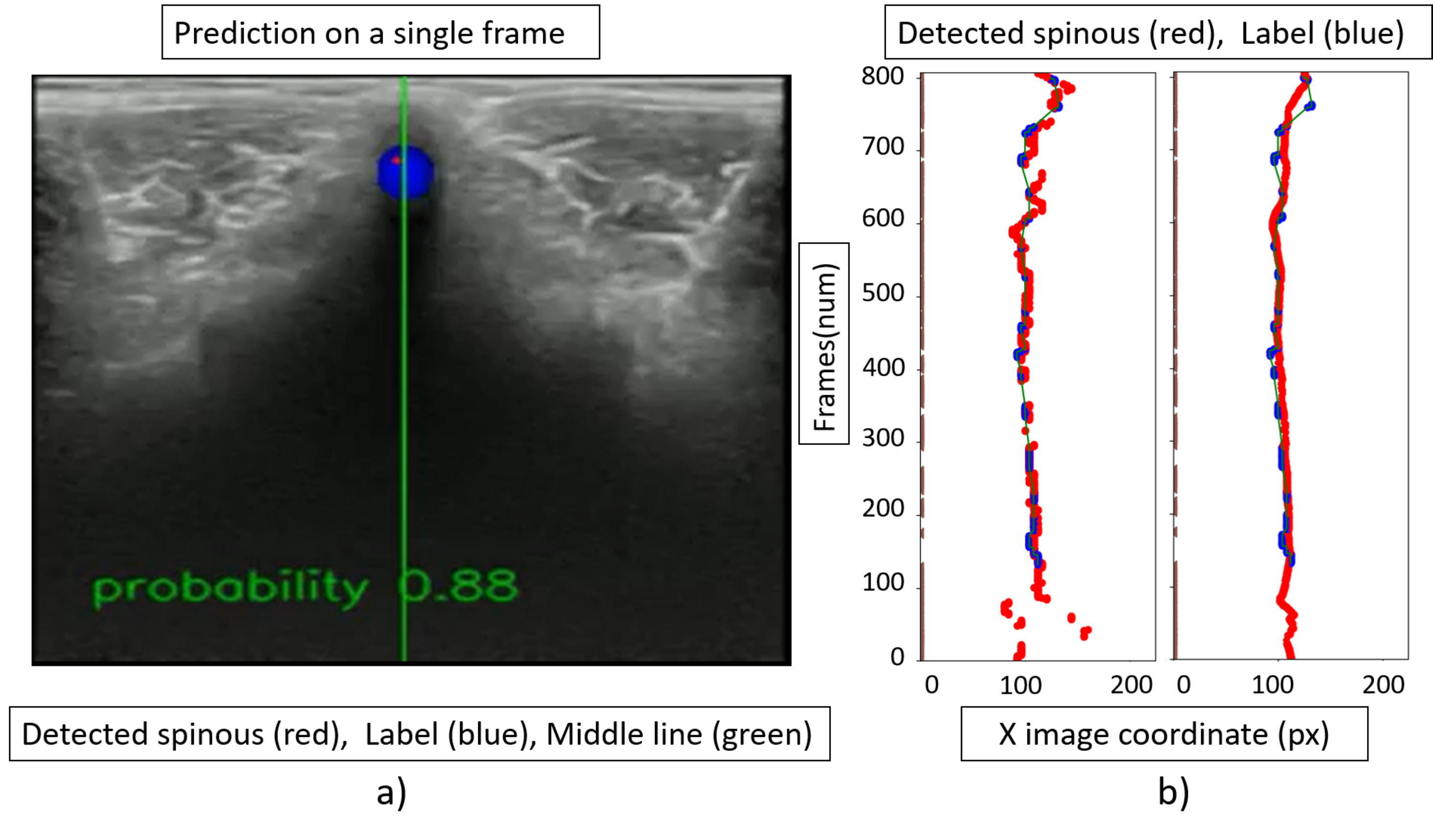}
    \caption[Spinous process detected path]{The results of the spinous detection on a single frame (a) and the resulted trajectory (b) LEFT. Real-time detection on a sweep (a sequence of US images obtained from human scan). Blue --- labels, red --- detected points with threshold of $50\%$. RIGHT. The resulted Kalman computed path.}
    \label{fig:detection_on_sweep_offline_results}
\end{figure}

Using the obtained model on a sequence of ultrasound images obtained from one human subject (referred to as ultrasound sweep) results in a sequence of points of predicted locations and the corresponding probability, which expresses the confidence of predicted location. The Fig. \ref{fig:detection_on_sweep_offline_results} b) LEFT demonstrates the spinous process presence in ultrasound sweep, obtained by manual scan. Red dots correspond to the predicted locations with the confidence of more than $50 \%$; blue dots are the correspondent target locations from manual labeling. For this exemplary ultrasound sweep, the mean accuracy of spinous process location predictions is $99.5 \%$ with a distance mean error of $0.8 mm$. 

However, predictions are made with high accuracy, there are still some noisy predictions, which present instability to the further robotic control, thus the Kalman filter was chosen for real-time prediction of the next frame spinous process location, taking new measurements as input only when the prediction accuracy is higher than the threshold of $50 \%$. The result of Kalman filtered path in US sweep is shown in Fig. \ref{fig:detection_on_sweep_offline_results} b) RIGHT. The noise covariance for process $Q=0.5$ and measurements $R=500$ were chosen experimentally. 
Before filtering, the mean accuracy on ultrasound sweeps of 5 test subjects is $89 \%$ and the mean distance error is $3.3 mm$.

\subsection{Real-time experiments of human examination with robotic-ultrasound system.}
\begin{figure}[]
    \centering
    \includegraphics[width=\linewidth]{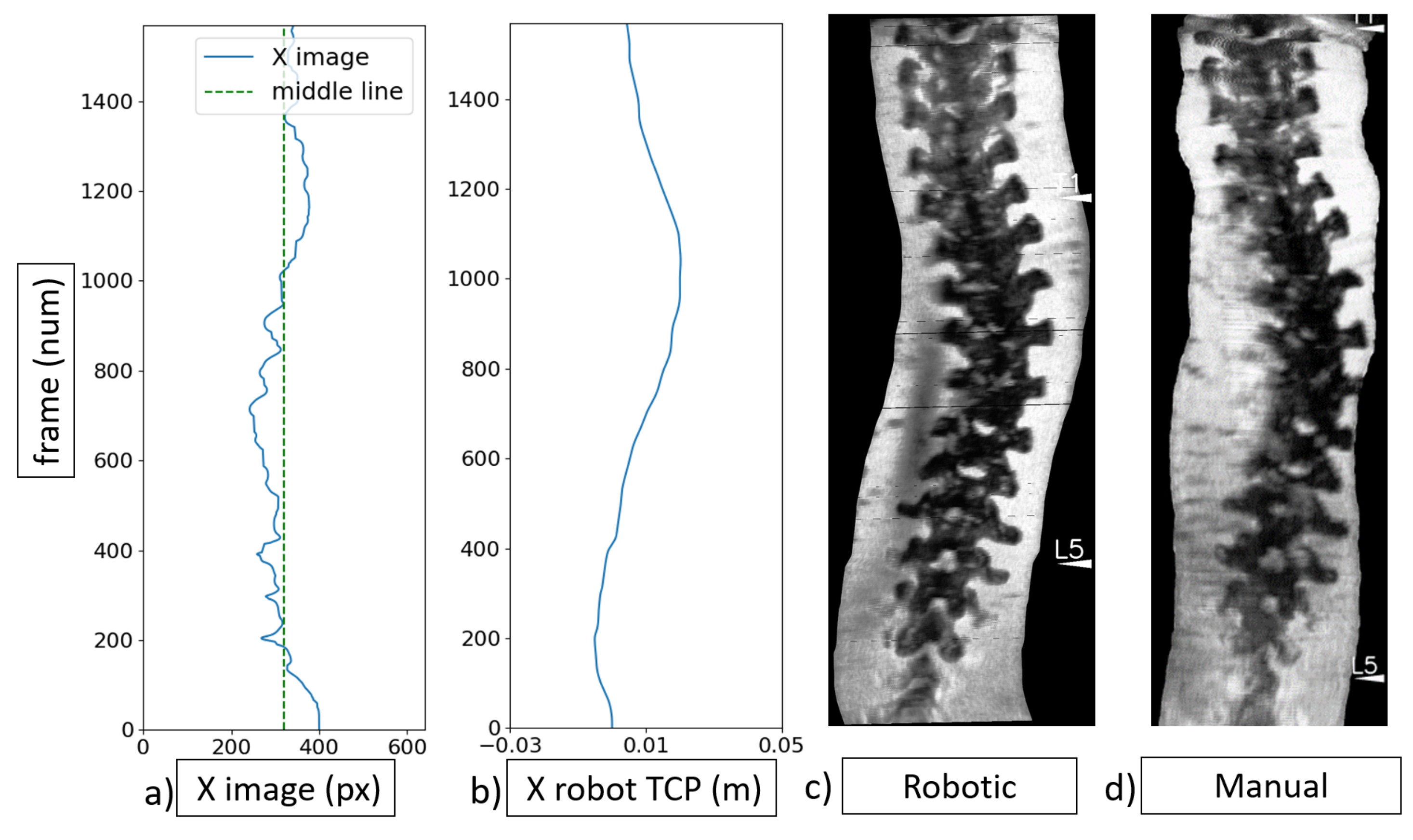}
    \caption[Spinous process tracking results on phantom]{Spinous process tracking results on phantom. a) The robot trajectory (m) on the axis with US-images control. b) Spinous process location at US image (pixels) frame by frame during the scan. c) Resulted reconstruction of the phantom's spine.}
    \label{fig:tracking_results_phantom}
\end{figure}

The proposed control approach was tested on a spinal phantom. Since the US images of the phantom are different from the real human spinal images, the separate model was obtained by training the same network as in Fig. \ref{fig:FCN_diagram} on 1,076 spinous process images obtained on the same phantom. Since the same phantom was used for training and testing, this test was done for the robotic control examination and not the model performance evaluation. Fig. \ref{fig:tracking_results_phantom} shows the results of phantom robotic scanning with speed of $v=0.004$, velocity output clipping of $v_{lim}=0.05$, Force $F_{ref}=7N$, PID control $K_p = 0.0004$ and tilt coefficient $k_{pitch}=0.07$. Kalman filter parameters were $Q = 0.5, R = 500$. The phantom scan: spinous process location mean $x_image = 322px$, with $STD=52 px$ $(6.5 mm)$.


\begin{figure}[]
    \centering
    \includegraphics[width=\linewidth]{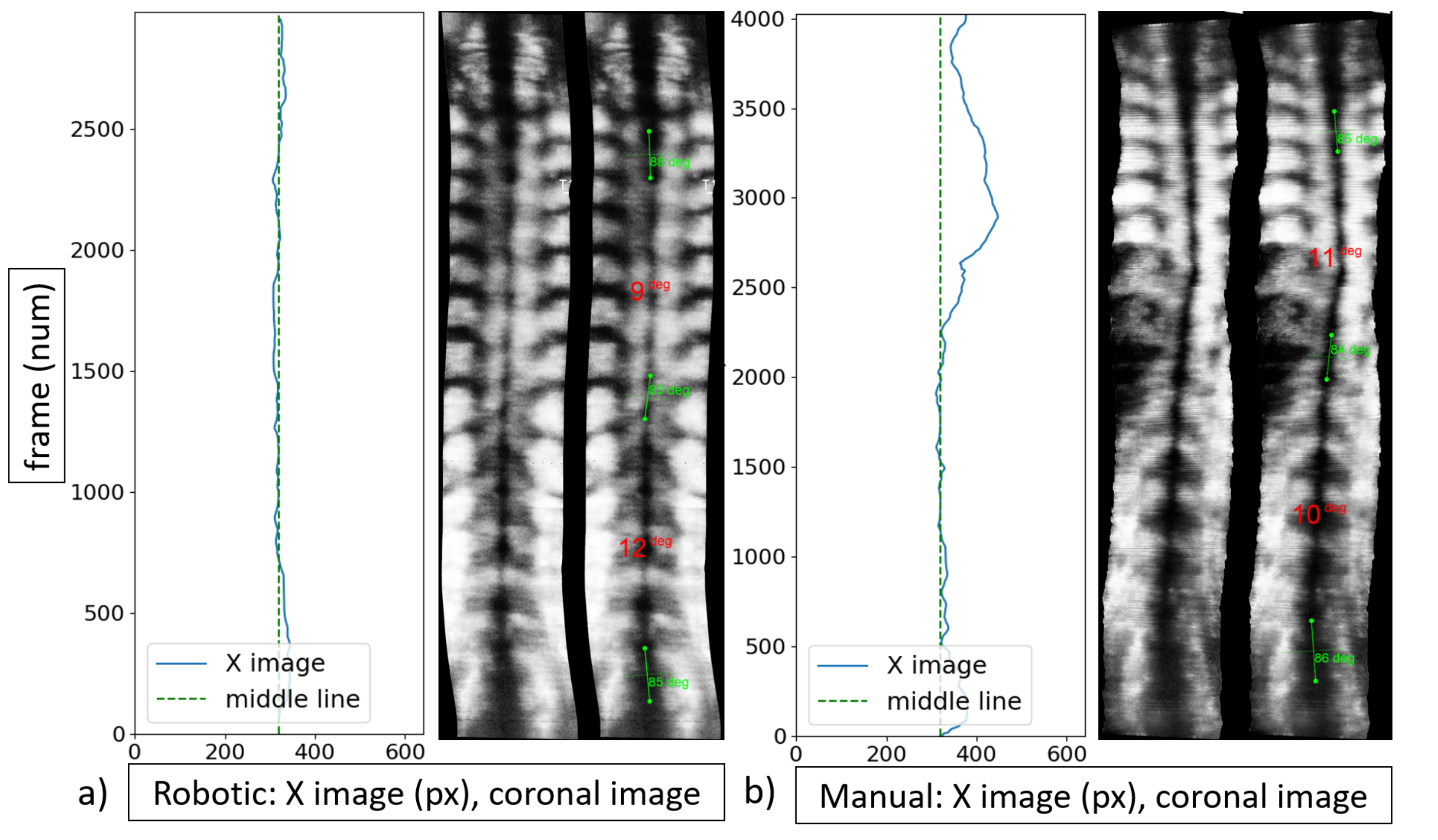}
    \caption[Spinous process tracking results on human]{Spinous process tracking results on human with mild Scoliosis. Kalman computed path (in pixels) frame by frame during the scan, where the green dash line is a middle line of the image. Resulted reconstruction of the human's spine from images collected during a) robotic scanning and b) manual scanning.}
    \label{fig:tracking_results_human_mild}
\end{figure}

Fig. \ref{fig:tracking_results_human_mild} shows the results of robotic scanning for human subject with mild severity of scoliosis (i.e. the deformity angle is less than 25 degrees). The plots display the Kalman computed paths for both robotic and manual scannings. The path for the manual scanning was computed after the manual procedure was performed, while for robotic approach the path was used for the navigation purpose. The scanning settings were the following: speed of $v=0.004$, velocity output clipping of $v_{lim}=0.002$, force setpoint $F=15N$ (the force range was chosen according to our previous work on spinal scanning \cite{Tirindelli2020}), PID control $K_p = 0.0003$, $K_d = K_i = 0.00003$ and tilt coefficient $K_{pitch} = 0.07$. Kalman filter  parameters for spinous process location output were $Q = 0.5, R = 500$. A mean deviation from the image center for the resulted spinous process path was calculated as $\delta x_{mean} = \frac{1}{N}\sum(x_i - w_{image}/2)$. For this case the $\delta x_{mean} = 7.8px$ $(1.0 mm)$ with $STD=6.5px$ for robotic approach and $\delta x_{mean} = 36.8px$ $(4.6 mm)$ with $STD=36.9px$ for manual scanning.

\begin{figure}[]
    \centering
    \includegraphics[width=\linewidth]{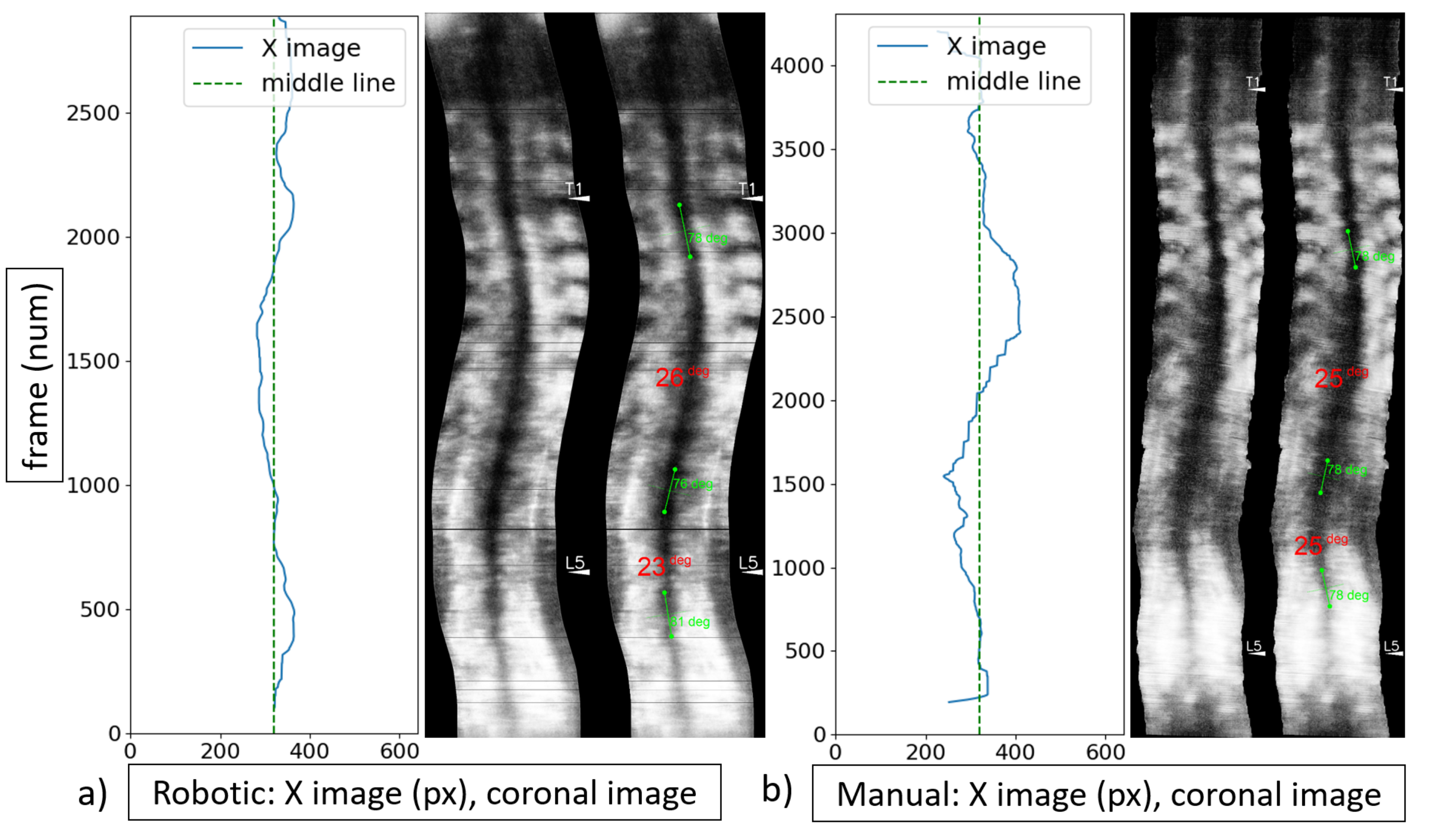}
    \caption[Spinous process tracking results on human]{Spinous process tracking results on human with moderate Scoliosis. Kalman computed path (in pixels) frame by frame during the scan, where the green dash line is a middle line of the image. Resulted reconstruction of the human's spine from images collected during a) robotic scanning and b) manual scanning.}
    \label{fig:tracking_results_human}
\end{figure}

The other example of the human subject with moderate scoliosis (i.e. the deformity angle is more than 25 degrees) is displayed at Fig. \ref{fig:tracking_results_human}. For this case the $\delta x_{mean} = 22.6px$ $(2.8 mm)$ with $STD=14.7px$ for robotic approach and $\delta x_{mean} = 34.7px$ $(4.3 mm)$ with $STD=31.5px$ for manual scanning.

\subsection{Spinal regions classification for parameters adjustments}

For region detection network was trained to classify ultrasound images of the ultrasound scan into four classes: Intervertebral gap, Lumbar spinous process, Sacrum, Thoracic spinous process. Together with class prediction, the network outputs the heatmap, where the maximum intensity corresponds to the spinous process location with a certain probability (the magnitude of the maximum intensity). The network was trained on the dataset, consisting of the spinous process, sacrum, gap ultrasound images. A total of 15 subjects (18,939 images) were used for training. The best model was achieved on 60 epoch with $LR = 0.001$, batch size of 12, Adam \cite{kingma2014adam} optimizer was used with learning rate decay of 0.5 every 20 epochs.
The weight loss for the heatmap loss was $C=1500$ and reflected the approximate magnitude ratios between classification head loss and deconvolutional head loss. The resulting 4-class model's validation accuracy on the test set of 5 subjects (6,835 images) is $84 \%$ for the classification task and $92 \%$ for the localization task.

\begin{figure*}[]
    \centering
    \includegraphics[width=\linewidth]{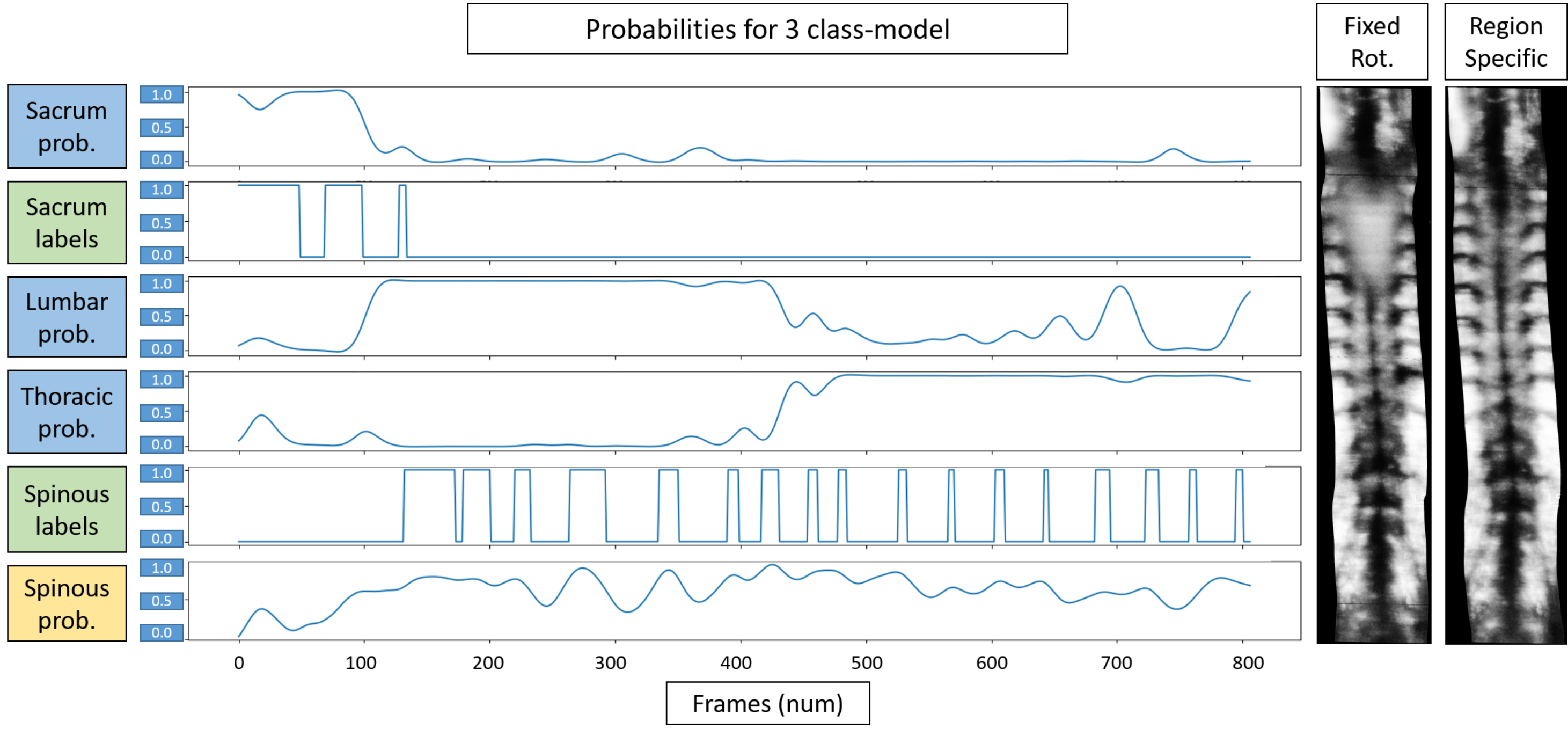}
    \caption[Spinal regions network output for 2-head network with classification head of 3 classes]{Example scan of human subject. Spinal regions network output for 2-head network with classification head of 3 classes. The resulted spinal coronal images with fixed rotation (LEFT) and with region specific rotation (RIGHT).}
    \label{fig:region_spine_result_3class}
\end{figure*}

The output of the 3-class multitask model on one of the test subjects is presented in Fig. \ref{fig:region_spine_result_3class}. The classes were: ``Sacrum", ``Lumbar spine", ``Thoracic spine". The images of only spinous process were used, without intervertebral gap images to imitate approach in Fig. \ref{fig:FCN_diagram}. The probabilities are presented in a range of 0 to 1, where 1 is a $100\%$ classification confidence. The Spinous probability results from the localization part of the network output, while other probabilities are the result of the classification output. 
The resulting 3-class model's validation accuracy on the test set of 5 subjects (6,835 images) is $95 \%$ for the classification task and $97 \%$ for the localization task.


\section{Discussion and Conclusion}
The spinous process localization algorithm yields excellent accuracy on the test dataset - $97.8\%$. The method presents to be robust in defining the spinous process location with a mean distance error of 1.0mm across the test set. 
Kalman filter works well on filtering the outliers, wrongly detected as spinous process, and preventing the robot from sudden sideways movements. The filter also forms the continuous path, filling in the locations gaps where the predictions on intervertebral gaps give accuracy less than 0.5 ($50\%$). 

The result of phantom scanning shows that the mean spinous process location among ultrasound frames was in $x_{image} = 322$ pixels, which is similar to the center of the ultrasound image $x_{image} = 320$ pixels. Visually assessing the tracking results for phantom, in Fig. \ref{fig:tracking_results_phantom}, it is clear that the coronal image generated from the robotic scan has spine centered in the field of view of the probe, the spinal features are more prominent, and the edges of the image are smoother. Since phantom is an idealistic case of the spine, the navigation performance was further assessed on human subjects. 

The human tracking results show how well the robotic approach could maintain the spinous process in the middle of US frames for participants with mild and moderate scoliosis (Fig. \ref{fig:tracking_results_human_mild} and Fig. \ref{fig:tracking_results_human}). The mean deviation from the image center of the detected spinous process during the scanning was between 1.0 mm for mild and 2.8 mm for the moderate case. The more significant deviation for the moderate case is explained by the greater angle of spinal deformity, where the robot takes longer time to center the spinous process in the field of view. Visual assessment of the image quality by the sonographer experts states that the robotic image for both cases looks smoother, the spinal features are more distinguishable. The last comment can be explained by the optimal force distribution along the spine. The robot applies a constant force, which is also generally slightly higher than what human does. 

The region pitch rotation adjustment turned to be an essential improvement for the system. Due to the probe's shape, the higher rotations cannot be used if the subject has significant lordosis at the lumbar region. Then lower rotation gain should be used $r_x$. However, if the same subject has a highly curved upper back, the lower rotations will not be sufficient, and the probe will lose contact with the subject's back. Thus, region-specific pitch rotation can significantly improve the quality of resulted scanning images.

The spinal region's network with or without ``intervertebral gaps" both perform well according to accuracy; the network without gaps was chosen since it has an output more straightforward for the lumbar and thoracic region splitting. The robotic experiment with region-specific pitch rotation clearly shows the improvement in the resulted coronal image, Fig. \ref{fig:region_spine_result_3class}. The robot could smoothly follow the spinal curvature, maintaining tight contact with the subject's skin, avoiding the air gap in the spine's thoracic (upper) region.

There are several limitations of the current work. There was no external visual feedback, which could be helpful when the subject has a narrow space between back bones (scapula) and the additional roll rotation is needed. Currently, there is no roll rotation designed in this study, which could allow aligning probe with the direction of the SP path. There were no severe scoliosis patients involved in this study, where due to the extreme spinal deformity, the spinous process is not present in the transverse ultrasound images for some vertebrae.
The project's next step is in progress: the system validation and reliability assessment in the subject's trials. It is vital to determine user cases where robotic scanning is practical or only the manual can be used, opening the path for future system improvements. 


\bibliographystyle{IEEEtran}
\bibliography{IEEEabrv,root.bib}

\end{document}